# MontiCore: A Framework for the Development of Textual Domain Specific Languages

Hans Grönniger    Holger Krahn

Bernhard Rumpe    Martin Schindler    Steven Völkel
Software Systems Engineering
TU Braunschweig, Germany

## ABSTRACT

In this paper we demonstrate a framework for efficient development of textual domain specific languages and supporting tools. We use a redundance-free and compact definition of a readable concrete syntax and a comprehensible abstract syntax as both representations significantly overlap in their structure. To further improve the usability of the abstract syntax this definition format integrates additional concepts like associations and inheritance into the well-understood grammar-based approach. Modularity concepts like language inheritance and embedding are used to simplify the development of languages based on already existing ones. In addition, the generation of editors and a template approach for code generation is explained.

## 1. INTRODUCTION

General purpose programming languages (GPLs) allow developers to create software systems efficiently. However, there is often a conceptual gap between the problem domain of an application and the used programming languages since GPLs typically do not contain domain abstractions. Domain specific languages (DSLs) provide a way to close this gap by enabling domain experts to describe a solution using familiar concepts and e.g., automatically map this description to an executable form. Experiments and case studies indicate that for certain domains major improvements can be achieved (see, e.g., [9]). Especially the evolution of systems with frequently changing requirements is easier, as DSLs are usually more concise and therefore smaller artifacts have to be adapted. In addition, DSLs allow a clear separation between the conceptual solution which is described with DSL code and technical aspects that are added by the code generator.

The MontiCore framework [6] can be used for agile development of simple as well as more complex textual DSLs. It uses a grammar-based language definition which is extended by several concepts to express associations, compositions, and inheritance directly in the language definition itself. Despite this concise language definition, the MontiCore framework supports the creation of a domain specific code generators by providing well-tested and practically approved solutions for re-occurring tasks.

## 2. LANGUAGE DEFINITION

MontiCore uses a single source for defining concrete as well as abstract syntax of a DSL. The abstract syntax derived from a grammar in a first step provides a tree. For an efficient navigation, derivation of typing information etc., we allow to specify additional associations, compositions and inheritance directly in the language definition. Tool support helps to automatically establish respectively calculate the links between the objects involved in an association. A MontiCore language definition is mapped to an object-oriented programming language where each nonterminal is mapped to a class with strongly typed attributes induced by the defining production. A parser is generated to create instances of the abstract syntax from a textual representation.

Figure 1 shows a grammar and the abstract syntax for finite automata, where an automaton starts with the keyword `automaton` and consists of an unbounded number of states and transitions. States can be initial or final and transition are activated by an input which leads to a state change. Each production is mapped to a class in the abstract syntax, each non-terminal is mapped to a composition relationship whereas references to terminals are implemented as attributes. In line 16-19 additional associations are added, whereas in lines 21-24 the concept `simplereference` is used to describe which instances are linked to each other. A more detailed description of the mapping can be found in [5].

Grammar inheritance as provided by MontiCore allows deriving new language variants from an existing language definition by specifying the delta only. The productions are added to the super-grammar and are used for parsing accordingly.

The concept of language embedding is realized by marking non-teminals as `external`. These external non-terminals (also called holes) can be filled by other grammars. Therefore, existing language fragments can be combined and reused in different settings. One main feature of MontiCore is that the fragments can be compiled independently and combined in compiled form. The advantage of this approach is that the fragments can be defined without knowledge about a concrete fragment combination. Furthermore, possible ambiguities are analyzed while designing the fragments and are therefore avoided at combination time. In addition the fragments can be deployed as compiled byte code which serves as a mechanism to protect the intellectual property of designing a parser.

```
                ─────── MontiCore-Grammar ───────
1  grammar Automaton {
2
3    Automaton =
4       "automaton" name:IDENT
5       "{" (State | Transition)* "}";
6
7    State =
8       "state" name:IDENT
9       (("<<" initial:["initial"] ">>" ) |
10      ("<<" final:["final"] ">>" ))* ";";
11
12   Transition =
13      from:IDENT "-" activate:IDENT ">"
14      to:IDENT ";";
15
16   associations {
17     Transition.toState *   <-> 1 State.ingoing;
18     Transition.fromState * <-> 1 State.outgoing;
19   }
20
21   concept simplereference {
22     toState: Transition.to -> State.name;
23     fromState: Transition.from -> State.name;
24   }
```

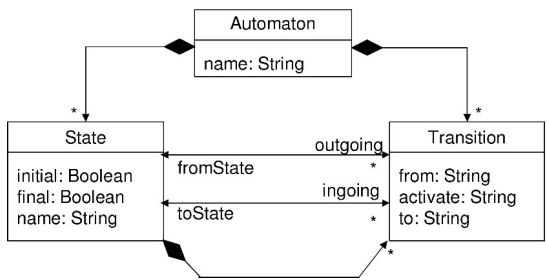

**Figure 1: Grammar and abstract syntax**

## 3. EDITOR AND CODE GENERATION

The editor environment for a newly defined language is a far too often neglected labor intensive task that does not directly contribute to the project, but only helps the developers to become more efficient later on. Therefore, a comfortable editor is an important success factor for a DSL if a new, and for other developers unknown DSL has to compete with a general purpose language like Java. For GPLs the existing tools are usually mature and support the user with a sophisticated user interface. DSLs therefore must be supplemented by a similar development environment.

We have chosen the Eclipse platform as a target for editor generation. The platform supports the user by a full-functional Java-IDE including among others an incremental compiler and the user is supported by various build and version management tools which are essential for efficient software development. A more detailed description can be found in [4].

The development of domain specific model analyzers and code generators is necessary to make a DSL useful. A code generator uses well-formed instances of the language that conform to the abstract syntax and context conditions and transforms them to executable code. MontiCore does not restrict the user to a certain form of code generation, as both a template engine or a visitor-base methods can be used. The development is supported by standard means to create files, process error messages, loading of related models, and structuring the processing by workflows which allows a developer to focus on the essentials of the generation process and rely on well-tested and practically approved solutions. In addition, MontiCore offers a template engine that supports template refactoring by a standard Java refactoring engine [3]. This simplifies the co-evolution of templates and runtime environment and therefore enables agile development of DSLs.

## 4. APPLICATIONS

MontiCore has been used for a number of applications for research and teaching purposes within the Institute for Software Systems Engineering. It allows the agile development of languages and simplifies experimental language design by not requiring additional libraries like EMF-based tools do.

The following list shows an excerpt of the realized projects.

- Autosar[1] is a component-based approach for modeling and implementing automotive software. In [2] it is shown how Autosar XSD files can be semi-automatically converted to a MontiCore grammar in order to describe the component interface directly next to the C-source code without the use of extra tools.

- A Java 5.0 compatible grammar is developed with the MontiCore framework and is commonly used as an action language for models.

- A subset of the UML [8, 7] which can be used as a programming language is described as a DSL in order to realize an UML tool supporting agile model-based development.

- The MontiCore framework is developed using its own infrastructure in a bootstrapping process.